\documentclass[prb,aps,twocolumn,reprint,showpacs,preprintnumbers,amsmath,amssymb,sort&compress,superscriptaddress,floatfix]{revtex4-1}
\usepackage{amsbsy,bm}
\usepackage[normalem]{ulem}

\newcommand{\av}[1]{\langle #1 \rangle}

\usepackage{multirow}
\usepackage{graphicx}
\usepackage{dcolumn}
\usepackage{epsfig,epstopdf}
\usepackage{subfigure}
\usepackage{color}

\begin{document}


\title{Decoherence of a qubit due to a quantum fluctuator or to a classical telegraph noise}
\author{Henry J. Wold}%
\affiliation{Department of Physics, University of Oslo, PO Box 1048 Blindern, 0316 Oslo, Norway}

\author{H{\aa}kon Brox}%
\affiliation{Department of Physics, University of Oslo, PO Box 1048 Blindern, 0316 Oslo, Norway}

\author{Yuri M. Galperin}%
\affiliation{Department of Physics, University of Oslo, PO Box 1048 Blindern, 0316 Oslo, Norway}
\affiliation{Centre for Advanced Study, Drammensveien 78, Oslo, Norway 0271, Oslo, Norway}
\affiliation{A. F. Ioffe Physico-Technical Institute of Russian Academy of Sciences, 194021 St. Petersburg, Russia}

\author{Joakim Bergli}%
\affiliation{Department of Physics, University of Oslo, PO Box 1048 Blindern, 0316 Oslo, Norway}

\date{\today}

\begin{abstract}
We investigate the decoherence of a qubit coupled to either a quantum
two-level system (TLS) again coupled to an environment, or a classical
fluctuator modeled by random telegraph noise. In order to do this we
construct a model for the quantum TLS where we can adjust the
temperature of its environment, and the decoherence rate
independently. The model has a well-defined classical limit at any
temperature and this corresponds to the appropriate random telegraph
process, which is symmetric at high temperatures and becomes
asymmetric at low temperatures. We find that the difference in the
qubit decoherence rates predicted by the two models depends on the
ratio between the qubit-TLS coupling and the decoherence rate in the
pointer basis of the TLS. This is then the relevant parameter which
determines whether the TLS has to be treated quantum mechanically or
can be replaced by a classical telegraph process. We also compare the
mutual information between the qubit and the TLS in the classical and
quantum cases.
\end{abstract}

\pacs{03.65.Yz, 03.67.Lx, 03.67.Bg, 74.78.Na}
\maketitle

\newcommand{\eq}{\! = \!}
\newcommand{\keq}{\!\! = \!\!}
\newcommand{\kadd}{\! + \!}

\section{Introduction}
The interaction between a quantum system and its environments leads to
loss of quantum coherence, or decoherence, in the system.
Understanding decoherence is 
crucial for grasping
the boundary between quantum and classical physics.~\cite{zurekd,paz,zurek,schloss}
It is also essential
for testing theories 
describing quantum measurements.~\cite{legett,wezel,ghirardi,adler} 

From an engineering point of view, 
the decay of coherence in 
quantum bit devices (qubits) is the most important obstacle for constructing 
a working quantum computer.
Solid state qubits are leading candidates in the projects of designing
quantum circuits, where the coherence times of the qubits are required
to be sufficiently long to allow for manipulations and transfer of
information by logical gates.  The most important source of
decoherence in many realizations of solid state qubits are believed to
be bistable fluctuators --
 two level systems (TLSs), present as tunneling states in
the amorphous substrate~\cite{halperin,phillipsart} used to fabricate the qubit, or in the
tunneling junction in superconductor-based
devices.~\cite{paladino02,martinis04,martinis05,martinis10,tian,shnirman05,zgalperinmeso,zgalperinprl}

These TLSs are quantum mechanical systems  that are, 
in turn,
coupled to their own environments, 
which are conventionally considered as uncorrelated thermal baths.
Usually one does not worry about the fine details of the environment of the TLSs,
but rather uses simplified models. The most popular is
the Bloch-Redfield approach,~\cite{slichter} where the
environment is taken into account by 
introduction of
the relaxation and decoherence
rates of the TLSs. If the TLSs couple more strongly to their own
environment than to the qubit, they are usually treated
classically. This means that the dynamical description of the TLSs is
replaced by a classical dynamics of a
fluctuator, which switches randomly between its
two metastable states according to a random telegraph process
(RTP).~\cite{kogan,kirton} 
This approach is often referred to as the
spin-fluctuator model.~\cite{paladino02,zgalperinprl,bergli2006}  In many cases, however, the
decoherence of the qubit is determined by only a few fluctuators that
are more strongly coupled to the qubit than
others.~\cite{simmonds,astafiev,koch,Bergli2005b,clemens} In such
cases, one might question the validity of the classical model.  From a
practical point of view, it is therefore important to know when such a
simplified classical description can replace the full quantum
mechanical one. It is also of more fundamental interest in view of the
decoherence approach to the quantum-classical
transition.~\cite{zurekd,paz,zurek,schloss}

In this paper, we will develop a simple model allowing to
show when a quantum system can in practice be replaced by a classical one, in the
sense that interference effects can no longer be observed due to the
entanglement with the environment. 
However, we believe that this is
only a question of a system becoming \emph{in practice} classical, i.e., when we can use a classical
model to calculate a physical property of a quantum system. It does not shed any light on the real limitation of quantum
mechanics such as the measurement problem, where one discuss deviations from linear quantum mechanics, see Ref.~\onlinecite{legett} for a discussion.

Previously, the boundary between quantum and classical regime for the
TLS has been explored in a model where the qubit is coupled to an
impurity state, and an electron can tunnel between this state and an
electron reservoir (metal).~\cite{grishin,abel} The qubit dephasing
rate calculated in the quantum model was found to converge to the
classical result in the high-temperature limit. In the study by Abel
and Marquardt,~\cite{abel} a threshold for strong coupling between the
qubit and the TLS was defined by the onset of visibility oscillations
in the qubit as a function of the ratio between the coupling to the
qubit and the reservoir. The threshold for visibility oscillations was
found for higher values of the qubit coupling in the quantum model
compared to the classical model, the thresholds finally converge at
high $T/\gamma$, where $\gamma$ is the TLS-reservoir coupling.  Thus,
both in the decoherence rate and in the visibility oscillations the
classical limit is recovered at high temperature.  In this model, the
temperature plays a dual role: It affects both the energy relaxation
rate of the TLS, which maps to the switching rate of the RTP, and it
affects the dephasing rate of the TLS. The 
usefulness of separation of
the two effects is seen by the fact that it is perfectly
possible to consider finite-temperature classical fluctuators by using
an asymetric RTP.\cite{falci,jung} This is never obtained in any
limit of the model discussed in Refs.~\onlinecite{grishin,abel}.

The subsequent considerations are based on
the following qualitative picture: The dephasing of the qubit
is caused by the generation of entanglement between the qubit and the
environment. If the qubit and the TLS are strongly coupled, 
then they 
behave as a combined four-level quantum system and the quantum nature
of the TLS will be important. In such a situation one cannot replace
it by a classical RTP. On the other hand, if the TLS is sufficiently
strongly coupled to the environment, it means that 
the information about
its state is continuously transferred to the environment and this
prevents any quantum interference to take place. From this we can
guess that the relevant quantity 
determining whether the TLS can
be considered 
either classical or 
quantum is the ratio of the qubit-TLS coupling
(which determines the rate of entanglement generation between the qubit
and TLS) and the TLS dephasing rate.  

The goal of this paper is to study the applicability of the classical
model for qubit decoherence due to a TLS.
In order to achieve this we study a model where the dephasing rate of the TLS can be varied
independently of the temperature, so that the classical limit can be
taken at any temperature and correspond to the proper assymetric RTP.
We investigate the regime where the TLS is coupled
weakly to the qubit. By use of a model borrowed from the study of TLSs
in glasses, we compare the pure decoherence rate of the qubit subject
to either a quantum TLS, in turn
coupled to its environment, or a
classical fluctuatur, modeled by random telegraph noise. Our
model allows us to separate the effects of temperature, coupling to
the bath, and decoherence rate of the TLS. We find that the difference
in the qubit decoherence rate predicted by the quantum 
model and the
classical one
depends on the ratio, $\xi/\bar{\gamma}_2$, where
$\xi$ is the qubit-TLS coupling strength and $\bar{\gamma}_2$ is the
 decoherence rate of the TLS in the pointer basis.

\section{Model}
\label{model}

\subsection{Quantum model for the TLS}
We start by describing the quantum mechanical model for the TLS.  The
model we use for the TLS originates in  the study of tunneling states in
glasses, i.e., a particle, or a group of particles that can be
approximated by a single configuratiol coordinate in a double-well
potential.~\cite{phillips}  If the particle is charged, it may give
rise to a potential on the qubit that depends on its position in the
double-well.

Following Phillips,~\cite{phillips} the Hamiltonian for the coupled
qubit-TLS is split into the Hamiltonians $H_q$ for the qubit, $H_f$
for the TLS, $H_i$ for the qubit-TLS interaction, $H_e$ for the
environment and $H_{fe}$ for the TLS-environment interaction:
\begin{align}
\label{Ham}
H&=H_q+H_f+H_i+H_e+H_{fe} , \nonumber\\
H_q&=E_q\tau_z \quad 
H_f =(1/2)\left(\Delta\sigma_z+\Delta_0\sigma_x\right), \nonumber \\
H_i& =(1/2) \, \xi\tau_z\sigma_z
\end{align}
where the Pauli matrices $\tau_{\alpha}$, $\sigma_{\alpha}$ are
operators in the Hilbert spaces of the qubit and the TLS,
respectively.  

The energy splitting, $\Delta$, and the tunnel
amplitude, $\Delta_0$, can be calculated from the shape of the
double-well potential.~\cite{phillips} The energy of the qubit depends
on the position of the particle in the double-well (we will in the
following refer to the eigenstates of $\sigma_z$ as the position
basis), and the coupling strength is given by $\xi$. In this work, we will assume the simplified case where the
qubit does not directly interact with the environment and therefore
has no intrinsic dynamics in the absence of the TLS.
Furthermore, we consider  a model where the qubit is subject to pure dephasing $[H_q,H_i]=0$,
there is no energy relaxation of the qubit in this model
and the decoherence of the
qubit is therefore insensitive to the qubit energy splitting $E_q$.
When energy relaxation is present, coherent beatings
between the qubit and resonant fluctuators are observed.~\cite{lisenfeld,simmonds}
In this strong coupling regime, the fluctuator has to be treated as a quantum system.
Our present work concentrates solely on non-resonant fluctuators, which are typically modeled classically.

The double-well potential is in general perturbed by electromagnetic
and strain fields modifying the asymmetry energy $\Delta$, while
perturbations of the barrier height can usually be
ignored.~\cite{acanderson,phillipsbook,galperin1989} In our model we therefore
assume that the environment couples to the TLS in the position basis,
i.e., the eigenbasis of $\sigma_z$.  Rather than formally specifying
$H_e$ and $H_{fe}$ we consider two kinds of interaction between the
TLS and the external environment, resonant and non-resonant. Resonant
interaction, e.g., phonons with frequency close to the eigenfrequency
of the TLS, are responsible for direct transitions between the
eigenstates of the TLS, $\left|\psi_g\right\rangle$ and
$\left|\psi_e\right\rangle$.  We model this interaction by use of the
generalized measurement operators
defined for a small time step $\Delta t$ as\cite{nielsen} 

\begin{align}
M_1(\Delta t)&=\sqrt{\gamma_{\text{ab}}(T)\Delta t}\,I\otimes\sigma_x\left|\psi_g\right\rangle\left\langle\psi_g\right|, \nonumber\\
M_2(\Delta t) &=\sqrt{\gamma_{\text{em}}(T)\Delta t}\,I\otimes\sigma_x\left|\psi_e\right\rangle\left\langle\psi_e\right|, \nonumber\\
M_3(\Delta t)&=\sqrt{1-M_1^{\dagger}M_1-M_2^{\dagger}M_2} \, .
\label{Mops}
\end{align}
Here $I$ is the identity matrix in the Hilbert space of the qubit and the matrices $\sigma_x\left|\psi_{g (e)}\right\rangle\left\langle\psi_{g (e)}\right|$ `measures' whether the TLS is in the ground (excited) state, projects the TLS onto this state and flips it. The rates for absorbtion and emission are 
\begin{align}
\gamma_{\text{ab}}(T)&=\gamma_1N(E) = \frac{\gamma_1}{e^{E/T}-1}, \nonumber\\
\gamma_{\text{em}}(T)&=\gamma_1[N(E)+1]
         =\frac{\gamma_1e^{E/T}}{e^{E/T}-1}.
\end{align}
Here $T$ is the temperature, $N(E) = \left(e^{E/T}-1\right)^{-1}$ is
the Planck distribution, and $E=\sqrt{\Delta^2+\Delta_0^2}$ is the energy
splitting of the TLS.
The non-resonant interaction does not cause transitions between the
eigenstates of the TLS. However, we might assume that in general the
state of a phonon interacting with the TLS is perturbed by the
interaction, and that the perturbation depends on the position of the
system in the double-well. Schematically we can write
\begin{align}
|\psi_i \rangle|\phi_0^{\text{ph}}\rangle&\stackrel{t}{\rightarrow}|\psi_i \rangle |\phi_i^{\text{ph}}\rangle,
\label{dec}
\end{align}  
where $i\in\{0,1\}$ index the state of the TLS in the position basis,
$|\phi_0^{\text{ph}}\rangle$ is the initial state of the phonon
and $|\phi_i^{\text{ph}}\rangle$ is the state of the phonon
after the interaction, conditioned upon that the TLS was initially in
the state indexed by $i$. 
The interaction, Eq.~\eqref{dec}, results in
entanglement between the phonon and the TLS, reducing the coherence of
the latter.  The rate of decoherence due to non-resonant phonons
depends on the overlap element $\alpha=\langle
\phi_0^{\text{ph}}|\phi_1^{\text{ph}}\rangle$ and on the rate of phonons
interacting with the system. We model this interaction by the single
parameter $\gamma_2$, which is  responsible for the decay rate of the
off-diagonal density matrix elements of the TLS in the position basis.
\begin{figure}[htb]
  \includegraphics[width=0.55\columnwidth]{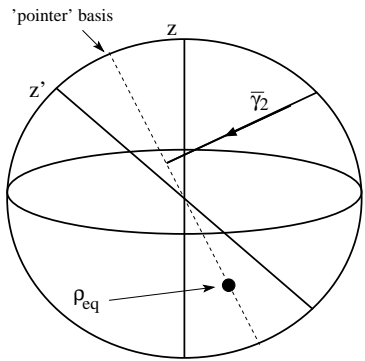}
 \caption{The Bloch-sphere for the TLS coupled to both non-resonant and resonant phonons. The non-resonant phonons are responsible for decay
perpendicular to the z-axis, the eigenbasis of $\sigma_z$, while the resonant phonons are responsible for relaxation parallell to the z'-axis, which is the eigenbasis of the TLS. We define the pointer basis by the basis in which the equilibrium density matrix $\rho_{eq}$ is diagonal. The rate of decay perpendicular to this axis is denoted by $\bar{\gamma}_2$.}
\label{coord}
\end{figure}

In this model the equilibrium density matrix of the TLS will not
necessarily lie along the $z$-axis of the Bloch sphere.
The equilibrium density matrix is determined by the rate $\gamma_2$ of
non-resonant phonons responsible for decay perpendicular to the $z$-axis on the Bloch-sphere
and by relaxation along the $z'$-axis due to resonant phonons in the eigenbasis of the TLS, 
at the rate $\gamma_1$ towards a level determined by $T$.
The situation is illustrated in Fig.~\ref{coord}.
We define the decoherence rate of the TLS, $\bar{\gamma}_2$,
by the rate at which the off-diagonal density matrix elements decay in
the basis where the density matrix is diagonal in equilibrium.

The time evolution in the quantum model is obtained by numerical integration of 
the von Neumann equation for the Hamiltonian given by Eq.~\eqref{Ham}, with two modifications.
We add a damping term $\gamma_2$ to our differential equation
\begin{align}
\dot{\rho}_{\alpha\alpha'}=i\left\langle\alpha\right|[\rho,H]\left|\alpha'\right\rangle-\Lambda_{\alpha\alpha'}\rho_{\alpha\alpha'},
\end{align}
where $\rho$ is the density matrix of the system composed of the qubit and the TLS and $\Lambda=\gamma_2I\otimes\sigma_x$ which
determines the decay of the off-diagonal density matrix elements of the TLS
in the eigenbasis of $\sigma_z$. In addition, the TLS absorb and emit phonons at the rates $\gamma_{\text{ab}}(T)$ and  $\gamma_{\text{em}}(T)$.
The absorption and emission of phonons is implemented as follows: for each timestep $\Delta t$ we make a transform to the eigenbasis
of the TLS
\begin{align}
\bar{\rho}=R(\theta)\rho R^{\dagger}(\theta),
\label{trans}
\end{align}
using the rotation matrix
$$R(\theta)=I\otimes\left( \begin{array}{cc}  \phantom{-}\cos \frac{ \theta}{2} &\sin \frac{\theta}{2} \\[0.02in]
-\sin\frac{\theta}{2} & \cos\frac{\theta}{2} \end{array} \right), \quad \theta\equiv \arctan\left(\frac{\Delta_0}{\Delta}\right). $$
The density matrix is then updated according to the rates of absorption and emission as
\begin{align}
\bar{\rho}' =M_1\bar{\rho}M_1^{\dagger}+M_2\bar{\rho}M_2^{\dagger}+M_3\bar{\rho}M_3^{\dagger},
\label{update}
\end{align}
before we make the inverse transform $\rho'=R^{\dagger}(\theta)\bar{\rho}'R(\theta)$, back to the position basis.
Here $\rho'$ is the density matrix after the (potential) interaction with the resonant phonons.

\subsection{Classical telegraph noise}
Pure dephasing of the qubit by a classical telegraph noise can be 
described by the interaction Hamiltonian 
\begin{align}
H_i&=(1/2)\xi(t)\tau_z,
\end{align}
 where $\xi(t)=\pm\xi$ is the position of the fluctuator at time $t$.
 For details on this model see, e.g., Ref.~\onlinecite{zbergli}
and references therein. The probability for the fluctuator to switch
from the state $\xi_-$ to $\xi_+$, and from $\xi_+$ to $\xi_-$ in the
interval $dt$ is given by $\Gamma_{-+}dt$ and $\Gamma_{+-}dt$,
respectively.  To describe finite temperature we will consider
the situation where the flipping rates $\Gamma_{-+}$ and $\Gamma_{+-}$
of the fluctuator are in general not identical, but the states are
symmetric $\xi_-=-\xi_+$. The situation with asymmetric switching
rates was previously studied in Refs.~\onlinecite{falci,jung}. The
equilibrium average is given by
\begin{equation}\label{RTPav}
\langle\xi\rangle = \xi ( p_{+}^{\text{eq}}- p_{-}^{\text{eq}})
  =\xi(\Gamma_{-+}-\Gamma_{+-})/\Gamma,
\end{equation}
where 
\begin{equation}\label{RTPGamma}
\Gamma=\Gamma_{-+}+\Gamma_{+-},
\end{equation}
 and $p_{{\pm}}(t)$ is the
probability for the fluctuator to be found in the state $\xi_{\pm}$, respectively.
The relaxation towards equilibrium is exponential with rate
 $\Gamma$.  The decoherence of the qubit is obtained by averaging
over the realizations and initial conditions of the noise process
$\xi(t)$. For a given realization of $\xi(t)$, the Schr{\"o}dinger equation yields
a superposition of the eigenstates of the qubit with a contribution to
the relative phase $\phi(t)=\int_0^t\xi(t')dt'$. Averaged over the realizations
of the stochastic process $\xi(t)$ we obtain the qubit coherence $D(t)=\left\langle e^{i\phi(t)}\right\rangle$.
  Here we will use the transfer matrix method developed by
Joynt~\emph{et al.},~\cite{joynt} where we obtain directly the
ensemble averaged Bloch-vector of the qubit.

The state of the qubit-fluctuator system can be stored in the six-dimensional vector
\begin{equation}
\vec{q}(t)=\vec{m}_+(t)\otimes\left(\begin{array}{c}1\\0\end{array}\right)p_+(t)+\vec{m}_-(t)\otimes\left(\begin{array}{c}0\\1\end{array}\right)p_-(t)
\end{equation}
where $\vec{m_{\pm}}$ is the Bloch vector of the qubit conditioned upon the state $\xi_{\pm}$ of the fluctuator. 
The propagator for $\vec{q}$ averaged over the individual realizations of the RTP 
can be expressed as $A(t)=e^{-Bt}$ where
$$ B=I_3\otimes V-i\frac{\xi}{2}L_z\otimes \upsilon_z,\quad V=\left(\begin{array}{cc}\phantom{-}  \Gamma_{+-}&-\Gamma_{-+}\\
-\Gamma_{+-}&\phantom{-} \Gamma_{-+}\end{array}\right),$$
while
$I_3$ and $L_z$ are generators of the SO$_3$ group and $\upsilon_Z$ is the Pauli matrix.
A direct advantage of this approach is that the qubit state conditioned upon whether the fluctuator is in the state $\xi_{\pm}$, $\rho_q^{\pm}$ 
follows directly from $\vec{q}$.

\section{Results}
In order to compare the decoherence of the qubit subject to either the
quantum TLS, or the classical telegraph noise, we 
calculate similar relaxation rates towards the equilibrium level in the two models.
First we choose a set of parameters, $\Delta$, $\Delta_0$, $\gamma_1$,
$\gamma_2$ and $T$ for the quantum model and prepare the TLS in the
initial state $|\psi_1\rangle$. At this preliminary stage
we are not interested in the qubit and consider the TLS and its
environment decoupled from the qubit. We compute numerically the
equilibrium occupation probabilities, $p_{0}^{\text{eq}}$ and $p_{1}^{\text{eq}}$, of
the TLS in the position basis, as well as the relaxation rate
$\Gamma$. Note that both the equilibrium occupations and the
relaxation rate are in general complicated functions of all the
parameters in our model. In this work, we always restrict ourselves to the
regime where the TLS is overdamped $\Delta,\Delta_0\ll \gamma_2$, i.e., the decoherence rate is sufficiently 
large such that coherent oscillations are not observed in the TLS. 
Parameter is needed.
Note also that since the states
$|\psi_i\rangle$ are not eigenstates of the Hamiltonian, the
occupation numbers $p_i^{\text{eq}}$ are not given by the Boltzmann weights
at the bath temperature.

The decoherence rate  is expressed through the
rates $\Gamma_{\pm \mp}$ and the equilibrium occupancy $\av{\xi}$ with the help of
Eqs. \eqref{RTPav} and \eqref{RTPGamma}.
The qubit decoherence rate is in general a sum over multiple
rates. For symmetric telegraph noise and pure dephasing the decay of coherence in the qubit $D(t)$ is given by~\cite{zbergli}
What is exact meaning of $D(t)$
\begin{equation}
D(t)=\frac{e^{-\Gamma t/2}}{2\mu}\left[(\mu+1)e^{\Gamma\mu t/2}+(\mu-1)e^{-\Gamma\mu t/2}\right],
\end{equation}
$\mu \equiv \sqrt{1 - (2\xi/\Gamma)^2}$,
but in the regime where the coupling to the
qubit is weak compared to the damping of the fluctuator, $\Gamma>\xi$,
the long-time behavior of the decoherence is strongly dominated by a
single rate, 
$$\Gamma_{q}^c = \Gamma(1-\mu)/2.$$
  
We finally compute the decoherence rate,
$\Gamma_q^{q}$, of the qubit when it is coupled to the same quantum
fluctuator from which we calculated the relaxation rate and
equilibrium occupations previously, but this time the initial state of the
TLS is the thermal equilibrium state. The decoherence rate of the qubit
is calculated by numerical simulation of the coupled qubit-fluctuator density matrix $\rho(t)$,
from which we can find the qubit density matrix by tracing out the degrees of freedom of the quantum fluctuator. From the qubit density matrix, $\rho^q(t)=\operatorname{Tr}_{f}[\rho(t)]$, we find the coherence $|\rho_{\uparrow\downarrow}^q(t)|$, where $\uparrow$ and $\downarrow$ denote the eigenstates of the qubit. Finally, $|\rho_{\uparrow\downarrow}^q(t)|$ is fitted to the exponential
function $e^{-\Gamma_q^{q}t}$.
 
The relative difference in the
decoherence rate of the qubit due to classical telegraph noise and the
quantum fluctuator is defined as
\begin{equation}
\delta\Gamma_q=( \Gamma_q^q-\Gamma_q^c)/\Gamma_q^c,  
\label{ratediff}
\end{equation}
where $\Gamma_q^{q}$ and $\Gamma_q^{c}$ are the decoherence rate of the
qubit subject to the quantum fluctuator and to the classical telegraph
noise, respectively.
\begin{figure}[htb]
  \includegraphics[width=0.99\columnwidth]{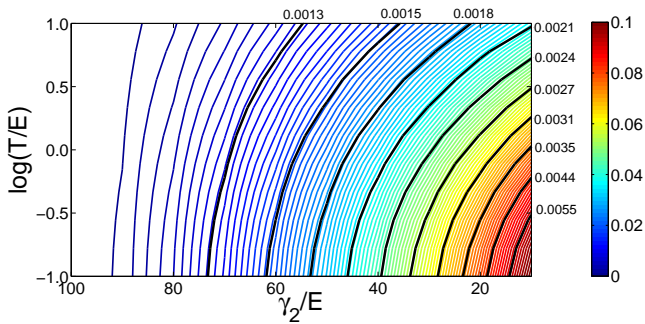}
 \caption{(Color online) 
Color lines -- contour plot of the
relative difference, $\delta\Gamma_q$, in the
decoherence rate of the qubit subject to either classical telegraph
noise, or a quantum fluctuator. In units of the TLS energy splitting $E$ the
parameters of the quantum fluctuator are $\Delta=\Delta_0=1/\sqrt{2}$ and
$\gamma_1=1.0$, the coupling to the qubit is $\xi=0.1$. 
Color coding for $\delta\Gamma_q$ is shown on the right.
 The relaxation rate to equilibrium along the $\sigma_z$-axis is the same
for both the quantum and the classical fluctuator.   
Contours where the
ratio $\xi/\bar{\gamma}_2$ is constant are plotted for comparison
(black lines).  }
\label{contour}
\end{figure}
This quantity  is presented
in Fig.~\ref{contour} as a function of the dephasing rate of the fluctuator, $\gamma_2$, and
temperature $T$.
We have
restricted ourselves to a parameter range where the TLS does not undergo
coherent oscillations. It is evident that the relative difference in
the qubit decoherence rate is small for strong decoherence of the TLS,
and for high temperatures. In this case we can safely use the simple
RTP model rather than the much more complicated quantum model.
Superimposed on the contours for $\delta\Gamma_q$ we have plotted
curves where the ratio $\xi/\bar{\gamma}_2$ is constant. 
We find that
the difference between the quantum and the classical fluctuator depends to
a very good accuracy  on the ratio $\xi/\bar{\gamma}_2$.
\begin{figure}[htb]
  \includegraphics[width=0.99\columnwidth]{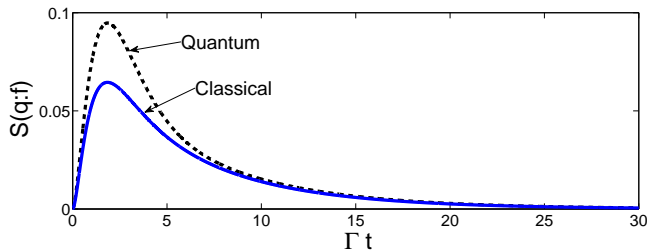}
 \caption{(Color online) 
Mutual information $S(q:f)$ for the qubit
coupled to the quantum TLS (black, dashed), and the qubit subject to
the classical spin-fluctuator (blue, solid). The mutual information is
larger when both systems are treated as quantum objects, due to
quantum entanglement between the two systems. In this simulation the
parameters, in units of $E$ are $\xi=0.1$, $\Delta=\Delta_0=1/\sqrt{2}$, $\gamma_1=1.0$, $\gamma_2=20$
and $E/T=1.0$. }
\label{mi}
\end{figure}

When the qubit is put in contact with the quantum TLS, the qubit and the
TLS will in general entangle due to their coupling. The mutual
information, the information about the state of one of the systems
that can be inferred by measuring the other, will for the quantum TLS
have an entanglement contribution in addition to the classical
correlation. 

The mutual information for the qubit-quantum TLS is defined
straightforwadly by the von Neumann entropy\cite{nielsen}
\begin{align}
S(q:f)&=S(\rho_q)+S(\rho_{f})-S(\rho_{qf}),
\label{eq:mi}
\end{align}
where $\rho_q$, $\rho_q$ and $\rho_{qf}$ are the density matrices of
the qubit, the TLS and the composite system, respectively.  When we
treat the qubit subject to a classical telegraph noise, we introduce
quantum states $|\pm \rangle$ corresponding to the states
$\xi_{\pm}$ of the RTP and use the formula
\begin{align}
\rho_{qf}&=p_+\rho_q^+\rho_{f+}+p_-\rho_q^-\rho_{f-}.
\end{align}
Here $p_\pm$ is the probability for the telegraph process to be
found in the state $\xi_{\pm}$, $\rho_q^\pm$ is the density
matrix of the qubit conditioned upon that the telegraph process is in the state
$\xi_{\pm}$ and
$\rho_{f\pm}=|\pm\rangle\langle\pm|$.

The time evolution of the mutual information for a qubit
coupled either to the quantum TLS or the classical fluctuator is shown
in Fig.~\ref{mi}. The entanglement between the two systems builds up
at a rate given by the coupling $\xi$, but is lost to the environment
at a rate given by the decoherence rate of the TLS,
$\bar{\gamma}_2$. The increased information about the qubit encoded in
the quantum TLS, compared to the classical fluctuator increases the
transfer of entropy to the environment, thus increasing the
decoherence rate of the qubit. This effect might explain the positive
$\delta\Gamma_q$ found for low values of $T$ and $\gamma_2$.

Experimentally, since the composite density matrix $\rho_{qf}$ is required, the mutual information can only be extracted in the 
case where one has access to measurement on both the qubit and the fluctuator simultaneously.
Since the fluctuator by definition is a system of the environment outside our control, this cannot be achieved.
However, the mutual information could potentially be studied in two coupled qubits,
where one of the qubits are subject to controlled noise and takes the role of the fluctuator.
Qubits subject to engineered noise under the control of the experimentalist has been realized in optically trapped $^9$Be$^+$ ions,~\cite{biercuk09}
where also the required quantum gates has already been implemented in a similar systems.~\cite{liebfried}

\section{Discussion}\label{discussion}

In general, the dynamics of a quantum TLS in an environment 
depends on three parameters; the relaxation rate $\gamma_1$, the
dephasing rate $\gamma_2$ and the temperature $T$  determining
the equilibrium occupations. In this paper, we use a model where the
processes responsible for pure dephasing couple to the position basis,
while the relaxation processes take place in the eigenbasis of the
TLS. This model was used in order to study the relevance of 
the classical RTP model for description of decoherence of a
qubit. 
If the
interaction responsible for pure dephasing processes (characterized by  $\gamma_2$) is
diagonal in the eigenbasis of the TLS, i.e., $\Delta_0=0$, it will not have any effect on
the decoherence rate of the qubit, as long as the qubit couples weakly
to the TLS $\xi/\gamma_1\ll 1$, and the TLS is prepared in the thermal
equilibrium state. The TLS will in this case \emph{always behave as a
classical fluctuator}, and can therefore straightforwardly be modeled
by the classical telegraph noise. 

In general, the difference in decoherence rate $\delta\Gamma$ depends
on the ratio $\Delta_0/\Delta$ as well as $\xi/\bar{\gamma}_2$, 
where the contours of constant $\xi/\bar{\gamma}_2$ in the $\ln T$ 
versus $\gamma_2$ plot, match those of constant $\delta\Gamma$ for all
values of  $\Delta_0/\Delta$.

Furthermore, we notice that our results 
do not 
tell us that it is, in principle, not possible to construct a classical telegraph model
providing
the same decoherence rate for the qubit as the quantum TLS, even
in the regime where the deviation $\delta\Gamma_q$ between the two
models are large according to Fig.~\ref{contour}. 
We show that the decoherence rate of the qubit differ in the two models
in the case where the relaxation rates of the classical and quantum fluctuator are identical.
To the best of our knowledge, there
exist no general relationship between the quantum 
TLS model and the classical 
spin-fluctuator model. 
Therefore, one should be careful in
applying the classical
telegraph model unless one expects the decoherence rates of the fluctuators
 to be much larger than the qubit-fluctuator coupling.
$\xi/\bar{\gamma}_2\ll 1$. However, in systems such as glasses this
inequality is usually expected to hold, and the TLS can be treated
effectively as a classical fluctuator,~\cite{phillips} with an
exception if the system is subject to an external AC field.~\cite{brox}

The pointer states of a quantum system are defined as the pure states
that are the least affected by environmental
decoherence.~\cite{zurekd,zurek} It is generally believed that when
the dynamics of the system is dominated by the interaction with the
environment, the pointer states are the eigenstates of the interaction
Hamiltonaian.~\cite{zurekd} On the other hand, when the system is
weakly coupled to the environment, the pointer states are assumed to
be the eigenstates of the isolated system.~\cite{paz} Our model can be
considered to interpolate between the two extremes. If we define the
pointer basis as the basis where the Bloch-vector of the system lies
along the $z$-axis in equilibrium, the decoherence rate
$\bar{\gamma}_2$ of the system is the rate of decay
of the off-diagonal elements of the density matrix in this basis. 

In conclusion, we have constructed a model for the quantum TLS where we
can study its effect on the qubit as a function of both temperature
and the decoherence rate of the TLS due to its interaction with the
environment. We have compared the decoherence rate of the qubit found
in this model, and in the widely used classical telegraph noise
model. We find that the difference in the qubit decoherence rates
depends on the ratio $\xi/\bar{\gamma}_2$ between the strength of the
qubit-TLS coupling and decoherence rate of the TLS in the pointer basis. In the
limit $\xi/\bar{\gamma}_2\ll 1$, the TLS behaves essentially
classically and the qubit decoherence rate can accurately be predicted
by the telegraph noise model.  

This work is part of the master project of one of the authors
(H. J. W.) and more details can be found in his thesis.~\cite{wold}


\begin{thebibliography}{43}
\expandafter\ifx\csname natexlab\endcsname\relax\def\natexlab#1{#1}\fi
\expandafter\ifx\csname bibnamefont\endcsname\relax
  \def\bibnamefont#1{#1}\fi
\expandafter\ifx\csname bibfnamefont\endcsname\relax
  \def\bibfnamefont#1{#1}\fi
\expandafter\ifx\csname citenamefont\endcsname\relax
  \def\citenamefont#1{#1}\fi
\expandafter\ifx\csname url\endcsname\relax
  \def\url#1{\texttt{#1}}\fi
\expandafter\ifx\csname urlprefix\endcsname\relax\def\urlprefix{URL }\fi
\providecommand{\bibinfo}[2]{#2}
\providecommand{\eprint}[2][]{\url{#2}}

\bibitem[{\citenamefont{Zurek}(1981)}]{zurekd}
\bibinfo{author}{\bibfnamefont{W.~H.} \bibnamefont{Zurek}},
  \bibinfo{journal}{Phys. Rev. D} \textbf{\bibinfo{volume}{24}},
  \bibinfo{pages}{1516} (\bibinfo{year}{1981}).

\bibitem[{\citenamefont{Paz and Zurek}(1999)}]{paz}
\bibinfo{author}{\bibfnamefont{J.~P.} \bibnamefont{Paz}} \bibnamefont{and}
  \bibinfo{author}{\bibfnamefont{W.~H.} \bibnamefont{Zurek}},
  \bibinfo{journal}{Phys. Rev. Lett.} \textbf{\bibinfo{volume}{82}},
  \bibinfo{pages}{5181} (\bibinfo{year}{1999}).

\bibitem[{\citenamefont{Zurek}(2003)}]{zurek}
\bibinfo{author}{\bibfnamefont{W.~H.} \bibnamefont{Zurek}},
  \bibinfo{journal}{Rev. Mod. Phys.} \textbf{\bibinfo{volume}{75}},
  \bibinfo{pages}{715} (\bibinfo{year}{2003}).

\bibitem[{\citenamefont{Schlosshauer}(2005)}]{schloss}
\bibinfo{author}{\bibfnamefont{M.}~\bibnamefont{Schlosshauer}},
  \bibinfo{journal}{Rev. Mod. Phys.} \textbf{\bibinfo{volume}{76}},
  \bibinfo{pages}{1267} (\bibinfo{year}{2005}).

\bibitem[{\citenamefont{Leggett}(2002)}]{legett}
\bibinfo{author}{\bibfnamefont{A.~J.} \bibnamefont{Leggett}},
  \bibinfo{journal}{J. Phys.: Cond. Mat.} \textbf{\bibinfo{volume}{14}},
  \bibinfo{pages}{R415} (\bibinfo{year}{2002}).

\bibitem[{\citenamefont{van Wezel and Oosterkamp}(2012)}]{wezel}
\bibinfo{author}{\bibfnamefont{J.}~\bibnamefont{van Wezel}} \bibnamefont{and}
  \bibinfo{author}{\bibfnamefont{T.~H.} \bibnamefont{Oosterkamp}},
  \bibinfo{journal}{Proc. R. Soc. A} \textbf{\bibinfo{volume}{468}},
  \bibinfo{pages}{35} (\bibinfo{year}{2012}).

\bibitem[{\citenamefont{Bassi and Ghirardi}(2003)}]{ghirardi}
\bibinfo{author}{\bibfnamefont{A.}~\bibnamefont{Bassi}} \bibnamefont{and}
  \bibinfo{author}{\bibfnamefont{G.}~\bibnamefont{Ghirardi}},
  \bibinfo{journal}{Phys. Rep.} \textbf{\bibinfo{volume}{379}},
  \bibinfo{pages}{257} (\bibinfo{year}{2003}).

\bibitem[{\citenamefont{Adler and Bassi}(2009)}]{adler}
\bibinfo{author}{\bibfnamefont{S.~L.} \bibnamefont{Adler}} \bibnamefont{and}
  \bibinfo{author}{\bibfnamefont{A.}~\bibnamefont{Bassi}},
  \bibinfo{journal}{Science} \textbf{\bibinfo{volume}{325}},
  \bibinfo{pages}{275} (\bibinfo{year}{2009}).

\bibitem[{\citenamefont{Anderson et~al.}(1972)\citenamefont{Anderson, I., and
  M.}}]{halperin}
\bibinfo{author}{\bibfnamefont{P.~W.} \bibnamefont{Anderson}},
  \bibinfo{author}{\bibfnamefont{H.~B.} \bibnamefont{I.}}, \bibnamefont{and}
  \bibinfo{author}{\bibfnamefont{V.~C.} \bibnamefont{M.}},
  \bibinfo{journal}{Phil. Mag.} \textbf{\bibinfo{volume}{25}},
  \bibinfo{pages}{1} (\bibinfo{year}{1972}).

\bibitem[{\citenamefont{Phillips}(1972)}]{phillipsart}
\bibinfo{author}{\bibfnamefont{W.~A.} \bibnamefont{Phillips}},
  \bibinfo{journal}{J. Low Temp. Phys.} \textbf{\bibinfo{volume}{7}},
  \bibinfo{pages}{351} (\bibinfo{year}{1972}).

\bibitem[{\citenamefont{Paladino et~al.}(2002)\citenamefont{Paladino, Faoro,
  Falci, and Fazio}}]{paladino02}
\bibinfo{author}{\bibfnamefont{E.}~\bibnamefont{Paladino}},
  \bibinfo{author}{\bibfnamefont{L.}~\bibnamefont{Faoro}},
  \bibinfo{author}{\bibfnamefont{G.}~\bibnamefont{Falci}}, \bibnamefont{and}
  \bibinfo{author}{\bibfnamefont{R.}~\bibnamefont{Fazio}},
  \bibinfo{journal}{PRL} \textbf{\bibinfo{volume}{88}}, \bibinfo{pages}{228304}
  (\bibinfo{year}{2002}).

\bibitem[{\citenamefont{Simmonds
  et~al.}(2004{\natexlab{a}})\citenamefont{Simmonds, Lang, Hite, Nam, Pappas,
  and Martinis}}]{martinis04}
\bibinfo{author}{\bibfnamefont{R.~W.} \bibnamefont{Simmonds}},
  \bibinfo{author}{\bibfnamefont{K.~M.} \bibnamefont{Lang}},
  \bibinfo{author}{\bibfnamefont{D.~A.} \bibnamefont{Hite}},
  \bibinfo{author}{\bibfnamefont{S.}~\bibnamefont{Nam}},
  \bibinfo{author}{\bibfnamefont{D.~P.} \bibnamefont{Pappas}},
  \bibnamefont{and} \bibinfo{author}{\bibfnamefont{J.~M.}
  \bibnamefont{Martinis}}, \bibinfo{journal}{Phys. Rev. Lett.}
  \textbf{\bibinfo{volume}{93}}, \bibinfo{pages}{077003}
  (\bibinfo{year}{2004}{\natexlab{a}}).

\bibitem[{\citenamefont{Martinis et~al.}(2005)\citenamefont{Martinis, Cooper,
  McDermott, Steffen, Ansmann, Osborn, Cicak, Oh, Pappas, Simmonds
  et~al.}}]{martinis05}
\bibinfo{author}{\bibfnamefont{J.~M.} \bibnamefont{Martinis}},
  \bibinfo{author}{\bibfnamefont{K.~B.} \bibnamefont{Cooper}},
  \bibinfo{author}{\bibfnamefont{R.}~\bibnamefont{McDermott}},
  \bibinfo{author}{\bibfnamefont{M.}~\bibnamefont{Steffen}},
  \bibinfo{author}{\bibfnamefont{M.}~\bibnamefont{Ansmann}},
  \bibinfo{author}{\bibfnamefont{K.~D.} \bibnamefont{Osborn}},
  \bibinfo{author}{\bibfnamefont{K.}~\bibnamefont{Cicak}},
  \bibinfo{author}{\bibfnamefont{S.}~\bibnamefont{Oh}},
  \bibinfo{author}{\bibfnamefont{D.~P.} \bibnamefont{Pappas}},
  \bibinfo{author}{\bibfnamefont{R.~W.} \bibnamefont{Simmonds}},
  \bibnamefont{et~al.}, \bibinfo{journal}{Phys. Rev. Lett.}
  \textbf{\bibinfo{volume}{95}}, \bibinfo{pages}{210503}
  (\bibinfo{year}{2005}).

\bibitem[{\citenamefont{Shalibo et~al.}(2010)\citenamefont{Shalibo, Rofe, Shwa,
  Zeides, Neeley, Martinis, and Katz}}]{martinis10}
\bibinfo{author}{\bibfnamefont{Y.}~\bibnamefont{Shalibo}},
  \bibinfo{author}{\bibfnamefont{Y.}~\bibnamefont{Rofe}},
  \bibinfo{author}{\bibfnamefont{D.}~\bibnamefont{Shwa}},
  \bibinfo{author}{\bibfnamefont{F.}~\bibnamefont{Zeides}},
  \bibinfo{author}{\bibfnamefont{M.}~\bibnamefont{Neeley}},
  \bibinfo{author}{\bibfnamefont{J.~M.} \bibnamefont{Martinis}},
  \bibnamefont{and} \bibinfo{author}{\bibfnamefont{N.}~\bibnamefont{Katz}},
  \bibinfo{journal}{Phys. Rev. Lett.} \textbf{\bibinfo{volume}{105}},
  \bibinfo{pages}{177001} (\bibinfo{year}{2010}).

\bibitem[{\citenamefont{Tian and Simmonds}(2007)}]{tian}
\bibinfo{author}{\bibfnamefont{L.}~\bibnamefont{Tian}} \bibnamefont{and}
  \bibinfo{author}{\bibfnamefont{R.~W.} \bibnamefont{Simmonds}},
  \bibinfo{journal}{Phys. Rev. Lett.} \textbf{\bibinfo{volume}{99}},
  \bibinfo{pages}{137002} (\bibinfo{year}{2007}).

\bibitem[{\citenamefont{Shnirman et~al.}(2005)\citenamefont{Shnirman, Sch\"on,
  Martin, and Makhlin}}]{shnirman05}
\bibinfo{author}{\bibfnamefont{A.}~\bibnamefont{Shnirman}},
  \bibinfo{author}{\bibfnamefont{G.}~\bibnamefont{Sch\"on}},
  \bibinfo{author}{\bibfnamefont{I.}~\bibnamefont{Martin}}, \bibnamefont{and}
  \bibinfo{author}{\bibfnamefont{Y.}~\bibnamefont{Makhlin}},
  \bibinfo{journal}{Phys. Rev. Lett.} \textbf{\bibinfo{volume}{94}},
  \bibinfo{pages}{127002} (\bibinfo{year}{2005}).

\bibitem[{\citenamefont{Galperin et~al.}(2004)\citenamefont{Galperin,
  Altshuler, and Shantsev}}]{zgalperinmeso}
\bibinfo{author}{\bibfnamefont{Y.~M.} \bibnamefont{Galperin}},
  \bibinfo{author}{\bibfnamefont{B.~L.} \bibnamefont{Altshuler}},
  \bibnamefont{and} \bibinfo{author}{\bibfnamefont{D.~V.}
  \bibnamefont{Shantsev}}, in \emph{\bibinfo{booktitle}{Fundamental Problems of
  Mesoscopic Physics}} (\bibinfo{publisher}{Springer Netherlands},
  \bibinfo{year}{2004}), vol. \bibinfo{volume}{154} of
  \emph{\bibinfo{series}{NATO Science Series}}, pp. \bibinfo{pages}{141--165}.

\bibitem[{\citenamefont{Galperin et~al.}(2006)\citenamefont{Galperin,
  Altshuler, Bergli, and Shantsev}}]{zgalperinprl}
\bibinfo{author}{\bibfnamefont{Y.~M.} \bibnamefont{Galperin}},
  \bibinfo{author}{\bibfnamefont{B.~L.} \bibnamefont{Altshuler}},
  \bibinfo{author}{\bibfnamefont{J.}~\bibnamefont{Bergli}}, \bibnamefont{and}
  \bibinfo{author}{\bibfnamefont{D.~V.} \bibnamefont{Shantsev}},
  \bibinfo{journal}{Phys. Rev. Lett.} \textbf{\bibinfo{volume}{96}},
  \bibinfo{pages}{097009} (\bibinfo{year}{2006}).

\bibitem[{\citenamefont{Schlichter}(1990)}]{slichter}
\bibinfo{author}{\bibfnamefont{C.~P.} \bibnamefont{Schlichter}},
  \emph{\bibinfo{title}{Principles of Magnetic Resonance}}
  (\bibinfo{publisher}{Springer-Verlag}, \bibinfo{year}{1990}).

\bibitem[{\citenamefont{Kogan}(1996)}]{kogan}
\bibinfo{author}{\bibfnamefont{S.}~\bibnamefont{Kogan}},
  \emph{\bibinfo{title}{Electronic noise and fluctuations in solids}}
  (\bibinfo{publisher}{Cambridge Univ. Press}, \bibinfo{year}{1996}).

\bibitem[{\citenamefont{Kirton and Uren}(1989)}]{kirton}
\bibinfo{author}{\bibfnamefont{M.}~\bibnamefont{Kirton}} \bibnamefont{and}
  \bibinfo{author}{\bibfnamefont{M.}~\bibnamefont{Uren}},
  \bibinfo{journal}{Adv. Phys.} \textbf{\bibinfo{volume}{38}},
  \bibinfo{pages}{367} (\bibinfo{year}{1989}).

\bibitem[{\citenamefont{Bergli et~al.}(2006)\citenamefont{Bergli, Galperin, and
  Altshuler}}]{bergli2006}
\bibinfo{author}{\bibfnamefont{J.}~\bibnamefont{Bergli}},
  \bibinfo{author}{\bibfnamefont{Y.~M.} \bibnamefont{Galperin}},
  \bibnamefont{and} \bibinfo{author}{\bibfnamefont{B.~L.}
  \bibnamefont{Altshuler}}, \bibinfo{journal}{Phys. Rev. B}
  \textbf{\bibinfo{volume}{74}}, \bibinfo{pages}{024509}
  (\bibinfo{year}{2006}).

\bibitem[{\citenamefont{Simmonds
  et~al.}(2004{\natexlab{b}})\citenamefont{Simmonds, Lang, Hite, Nam, Pappas,
  and Martinis}}]{simmonds}
\bibinfo{author}{\bibfnamefont{R.~W.} \bibnamefont{Simmonds}},
  \bibinfo{author}{\bibfnamefont{K.~M.} \bibnamefont{Lang}},
  \bibinfo{author}{\bibfnamefont{D.~A.} \bibnamefont{Hite}},
  \bibinfo{author}{\bibfnamefont{S.}~\bibnamefont{Nam}},
  \bibinfo{author}{\bibfnamefont{D.~P.} \bibnamefont{Pappas}},
  \bibnamefont{and} \bibinfo{author}{\bibfnamefont{J.~M.}
  \bibnamefont{Martinis}}, \bibinfo{journal}{Phys. Rev. Lett.}
  \textbf{\bibinfo{volume}{93}}, \bibinfo{pages}{077003}
  (\bibinfo{year}{2004}{\natexlab{b}}).

\bibitem[{\citenamefont{Astafiev et~al.}(2004)\citenamefont{Astafiev, Pashkin,
  Nakamura, Yamamoto, and Tsai}}]{astafiev}
\bibinfo{author}{\bibfnamefont{O.}~\bibnamefont{Astafiev}},
  \bibinfo{author}{\bibfnamefont{Y.~A.} \bibnamefont{Pashkin}},
  \bibinfo{author}{\bibfnamefont{Y.}~\bibnamefont{Nakamura}},
  \bibinfo{author}{\bibfnamefont{T.}~\bibnamefont{Yamamoto}}, \bibnamefont{and}
  \bibinfo{author}{\bibfnamefont{J.~S.} \bibnamefont{Tsai}},
  \bibinfo{journal}{Phys. Rev. Lett.} \textbf{\bibinfo{volume}{93}},
  \bibinfo{pages}{267007} (\bibinfo{year}{2004}).

\bibitem[{\citenamefont{Koch et~al.}(2007)\citenamefont{Koch, DiVincenzo, and
  Clarke}}]{koch}
\bibinfo{author}{\bibfnamefont{R.~H.} \bibnamefont{Koch}},
  \bibinfo{author}{\bibfnamefont{D.~P.} \bibnamefont{DiVincenzo}},
  \bibnamefont{and} \bibinfo{author}{\bibfnamefont{J.}~\bibnamefont{Clarke}},
  \bibinfo{journal}{Phys. Rev. Lett.} \textbf{\bibinfo{volume}{98}},
  \bibinfo{pages}{267003} (\bibinfo{year}{2007}).

\bibitem[{\citenamefont{Galperin et~al.}(2005)\citenamefont{Galperin, Shantsev,
  Bergli, and Altshuler}}]{Bergli2005b}
\bibinfo{author}{\bibfnamefont{Y.~M.} \bibnamefont{Galperin}},
  \bibinfo{author}{\bibfnamefont{D.~V.} \bibnamefont{Shantsev}},
  \bibinfo{author}{\bibfnamefont{J.}~\bibnamefont{Bergli}}, \bibnamefont{and}
  \bibinfo{author}{\bibfnamefont{B.~L.} \bibnamefont{Altshuler}},
  \bibinfo{journal}{Europhys. Lett.} \textbf{\bibinfo{volume}{71}},
  \bibinfo{pages}{21} (\bibinfo{year}{2005}).

\bibitem[{\citenamefont{M\"uller et~al.}(2009)\citenamefont{M\"uller, Shnirman,
  and Makhlin}}]{clemens}
\bibinfo{author}{\bibfnamefont{C.}~\bibnamefont{M\"uller}},
  \bibinfo{author}{\bibfnamefont{A.}~\bibnamefont{Shnirman}}, \bibnamefont{and}
  \bibinfo{author}{\bibfnamefont{Y.}~\bibnamefont{Makhlin}},
  \bibinfo{journal}{Phys. Rev. B} \textbf{\bibinfo{volume}{80}},
  \bibinfo{pages}{134517} (\bibinfo{year}{2009}).

\bibitem[{\citenamefont{Grishin et~al.}(2005)\citenamefont{Grishin, Yurkevich,
  and Lerner}}]{grishin}
\bibinfo{author}{\bibfnamefont{A.}~\bibnamefont{Grishin}},
  \bibinfo{author}{\bibfnamefont{I.~V.} \bibnamefont{Yurkevich}},
  \bibnamefont{and} \bibinfo{author}{\bibfnamefont{I.~V.}
  \bibnamefont{Lerner}}, \bibinfo{journal}{Phys. Rev. B}
  \textbf{\bibinfo{volume}{72}}, \bibinfo{pages}{060509}
  (\bibinfo{year}{2005}).

\bibitem[{\citenamefont{Abel and Marquardt}(2008)}]{abel}
\bibinfo{author}{\bibfnamefont{B.}~\bibnamefont{Abel}} \bibnamefont{and}
  \bibinfo{author}{\bibfnamefont{F.}~\bibnamefont{Marquardt}},
  \bibinfo{journal}{Phys. Rev. B} \textbf{\bibinfo{volume}{78}},
  \bibinfo{pages}{201302} (\bibinfo{year}{2008}).

\bibitem[{\citenamefont{Falci et~al.}(2003)\citenamefont{Falci, Paladino, and
  Fazio}}]{falci}
\bibinfo{author}{\bibfnamefont{G.}~\bibnamefont{Falci}},
  \bibinfo{author}{\bibfnamefont{E.}~\bibnamefont{Paladino}}, \bibnamefont{and}
  \bibinfo{author}{\bibfnamefont{R.}~\bibnamefont{Fazio}}, in
  \emph{\bibinfo{booktitle}{Quantum Phenomena in Mesoscopic Systems}}
  (\bibinfo{publisher}{IOS Press Amsterdam}, \bibinfo{year}{2003}), vol.
  \bibinfo{volume}{151} of \emph{\bibinfo{series}{Proceedings of the
  International School of Physics "Enrico Fermi"}}, pp.
  \bibinfo{pages}{173--198}.

\bibitem[{\citenamefont{Jung et~al.}(2002)\citenamefont{Jung, Barkai, and
  Silbey}}]{jung}
\bibinfo{author}{\bibfnamefont{Y.}~\bibnamefont{Jung}},
  \bibinfo{author}{\bibfnamefont{E.}~\bibnamefont{Barkai}}, \bibnamefont{and}
  \bibinfo{author}{\bibfnamefont{R.~J.} \bibnamefont{Silbey}},
  \bibinfo{journal}{Chem. Phys.} \textbf{\bibinfo{volume}{284}},
  \bibinfo{pages}{181 } (\bibinfo{year}{2002}).

\bibitem[{\citenamefont{Phillips}(1987)}]{phillips}
\bibinfo{author}{\bibfnamefont{W.~A.} \bibnamefont{Phillips}},
  \bibinfo{journal}{Rep. Prog. Phys.} \textbf{\bibinfo{volume}{50}},
  \bibinfo{pages}{1657} (\bibinfo{year}{1987}).

\bibitem[{\citenamefont{Lisenfeld et~al.}(2010)\citenamefont{Lisenfeld,
  M\"uller, Cole, Bushev, Lukashenko, Shnirman, and Ustinov}}]{lisenfeld}
\bibinfo{author}{\bibfnamefont{J.}~\bibnamefont{Lisenfeld}},
  \bibinfo{author}{\bibfnamefont{C.}~\bibnamefont{M\"uller}},
  \bibinfo{author}{\bibfnamefont{J.~H.} \bibnamefont{Cole}},
  \bibinfo{author}{\bibfnamefont{P.}~\bibnamefont{Bushev}},
  \bibinfo{author}{\bibfnamefont{A.}~\bibnamefont{Lukashenko}},
  \bibinfo{author}{\bibfnamefont{A.}~\bibnamefont{Shnirman}}, \bibnamefont{and}
  \bibinfo{author}{\bibfnamefont{A.~V.} \bibnamefont{Ustinov}},
  \bibinfo{journal}{Phys. Rev. Lett.} \textbf{\bibinfo{volume}{105}},
  \bibinfo{pages}{230504} (\bibinfo{year}{2010}).

\bibitem[{\citenamefont{Anderson}(1986)}]{acanderson}
\bibinfo{author}{\bibfnamefont{A.~C.} \bibnamefont{Anderson}},
  \bibinfo{journal}{Phys. Rev. B} \textbf{\bibinfo{volume}{34}},
  \bibinfo{pages}{1317} (\bibinfo{year}{1986}).

\bibitem[{\citenamefont{Phillips}(1981)}]{phillipsbook}
\bibinfo{author}{\bibfnamefont{W.~A.} \bibnamefont{Phillips}},
  \emph{\bibinfo{title}{Amorphous solids: low-temperature properties}},
  vol.~\bibinfo{volume}{24} (\bibinfo{publisher}{Berlin:Springer},
  \bibinfo{year}{1981}).

\bibitem[{\citenamefont{Galperin et~al.}(1989)\citenamefont{Galperin, Karpov,
  and Kozub}}]{galperin1989}
\bibinfo{author}{\bibfnamefont{Y.~M.} \bibnamefont{Galperin}},
  \bibinfo{author}{\bibfnamefont{V.~G.} \bibnamefont{Karpov}},
  \bibnamefont{and} \bibinfo{author}{\bibfnamefont{V.~I.} \bibnamefont{Kozub}},
  \bibinfo{journal}{Adv. Phys.} \textbf{\bibinfo{volume}{38}},
  \bibinfo{pages}{669} (\bibinfo{year}{1989}).

\bibitem[{\citenamefont{Nielsen and Chuang}(2000)}]{nielsen}
\bibinfo{author}{\bibfnamefont{M.~A.} \bibnamefont{Nielsen}} \bibnamefont{and}
  \bibinfo{author}{\bibfnamefont{I.~L.} \bibnamefont{Chuang}},
  \emph{\bibinfo{title}{Quantum Computation and Quantum Information}}
  (\bibinfo{publisher}{Cambridge University Press}, \bibinfo{year}{2000}).

\bibitem[{\citenamefont{Bergli et~al.}(2009)\citenamefont{Bergli, Galperin, and
  Altshuler}}]{zbergli}
\bibinfo{author}{\bibfnamefont{J.}~\bibnamefont{Bergli}},
  \bibinfo{author}{\bibfnamefont{Y.~M.} \bibnamefont{Galperin}},
  \bibnamefont{and} \bibinfo{author}{\bibfnamefont{B.~L.}
  \bibnamefont{Altshuler}}, \bibinfo{journal}{New J. Phys.}
  \textbf{\bibinfo{volume}{11}}, \bibinfo{pages}{025002}
  (\bibinfo{year}{2009}).

\bibitem[{\citenamefont{Cheng et~al.}(2008)\citenamefont{Cheng, Wang, and
  Joynt}}]{joynt}
\bibinfo{author}{\bibfnamefont{B.}~\bibnamefont{Cheng}},
  \bibinfo{author}{\bibfnamefont{Q.-H.} \bibnamefont{Wang}}, \bibnamefont{and}
  \bibinfo{author}{\bibfnamefont{R.}~\bibnamefont{Joynt}},
  \bibinfo{journal}{Phys. Rev. A} \textbf{\bibinfo{volume}{78}},
  \bibinfo{pages}{022313} (\bibinfo{year}{2008}).

\bibitem[{\citenamefont{Biercuk et~al.}(2009)\citenamefont{Biercuk, Uys,
  VanDevender, Shiga, Itano, and Bollinger}}]{biercuk09}
\bibinfo{author}{\bibfnamefont{M.~J.} \bibnamefont{Biercuk}},
  \bibinfo{author}{\bibfnamefont{H.}~\bibnamefont{Uys}},
  \bibinfo{author}{\bibfnamefont{A.~P.} \bibnamefont{VanDevender}},
  \bibinfo{author}{\bibfnamefont{N.}~\bibnamefont{Shiga}},
  \bibinfo{author}{\bibfnamefont{W.~M.} \bibnamefont{Itano}}, \bibnamefont{and}
  \bibinfo{author}{\bibfnamefont{J.~J.} \bibnamefont{Bollinger}},
  \bibinfo{journal}{Phys. Rev. A} \textbf{\bibinfo{volume}{79}},
  \bibinfo{pages}{062324} (\bibinfo{year}{2009}).

\bibitem[{\citenamefont{Leibfried et~al.}(2003)\citenamefont{Leibfried,
  DeMarco, Meyer, Lucas, Barrett, Britton, Itano, Jelenkovic, Langer, Rosenband
  et~al.}}]{liebfried}
\bibinfo{author}{\bibfnamefont{D.}~\bibnamefont{Leibfried}},
  \bibinfo{author}{\bibfnamefont{B.}~\bibnamefont{DeMarco}},
  \bibinfo{author}{\bibfnamefont{V.}~\bibnamefont{Meyer}},
  \bibinfo{author}{\bibfnamefont{D.}~\bibnamefont{Lucas}},
  \bibinfo{author}{\bibfnamefont{M.}~\bibnamefont{Barrett}},
  \bibinfo{author}{\bibfnamefont{J.}~\bibnamefont{Britton}},
  \bibinfo{author}{\bibfnamefont{W.~M.} \bibnamefont{Itano}},
  \bibinfo{author}{\bibfnamefont{B.}~\bibnamefont{Jelenkovic}},
  \bibinfo{author}{\bibfnamefont{C.}~\bibnamefont{Langer}},
  \bibinfo{author}{\bibfnamefont{T.}~\bibnamefont{Rosenband}},
  \bibnamefont{et~al.}, \bibinfo{journal}{Nature}
  \textbf{\bibinfo{volume}{422}}, \bibinfo{pages}{412} (\bibinfo{year}{2003}).

\bibitem[{\citenamefont{Brox et~al.}(2011)\citenamefont{Brox, Bergli, and
  Galperin}}]{brox}
\bibinfo{author}{\bibfnamefont{H.}~\bibnamefont{Brox}},
  \bibinfo{author}{\bibfnamefont{J.}~\bibnamefont{Bergli}}, \bibnamefont{and}
  \bibinfo{author}{\bibfnamefont{Y.~M.} \bibnamefont{Galperin}},
  \bibinfo{journal}{Phys. Rev. B} \textbf{\bibinfo{volume}{84}},
  \bibinfo{pages}{245416} (\bibinfo{year}{2011}).

\bibitem[{\citenamefont{Wold}(2011)}]{wold}
\bibinfo{author}{\bibfnamefont{H.~J.} \bibnamefont{Wold}}, Master's thesis,
  \bibinfo{school}{University of Oslo, Norway} (\bibinfo{year}{2011}),
  \bibinfo{note}{http://urn.nb.no/URN:NBN:no-28902}.

\end{thebibliography}
\end{document}